\documentclass[%
 reprint,
 amsmath,amssymb,
 aps,
]{revtex4-1}

\usepackage{graphicx}
\usepackage{dcolumn}
\usepackage{bm}

\usepackage{color}

\begin{document}

\preprint{APS/123-QED}

\title{Critical Filling Factor for the Formation of a Quantum Wigner Crystal \\Screened by a Nearby Layer}

\author{H.\ Deng, L.N.\ Pfeiffer, K.W.\ West, K.W.\ Baldwin, and M.\ Shayegan}

\affiliation{Department of Electrical Engineering, Princeton University}

\date{\today}

\begin{abstract}
One of the most fascinating ground states of an interacting electron system
is the so-called Wigner crystal where the electrons,
in order to minimize their repulsive Coulomb energy, form an ordered array.
Here we report measurements of the critical filling factor ($\nu_{C}$)
below which a magnetic-field-induced, quantum Wigner crystal
forms in a dilute, two-dimensional electron layer
when a second, high-density electron layer
is present in close proximity.
The data reveal that the Wigner crystal forms at a significantly smaller $\nu_{C}$
compared to the $\nu_{C}$ ($\simeq 0.20$) in single-layer two-dimensional electron systems.
The measured $\nu_{C}$ exhibits a strong dependence on the interlayer distance,
reflecting the interaction and screening from the adjacent, high-density layer.
\end{abstract}

\pacs{Valid PACS appear here}
\maketitle


When the Coulomb energy ($E_C$) in an interacting electron system
dominates over the kinetic energy,
it has long been expected that the system condenses into a Wigner crystal (WC)
where electrons order themselves in a periodic lattice \cite{Wigner.PR.46.1002}.
A \textit{classical} WC state was indeed realized
in a very dilute two-dimensional electron system (2DES)
confined to the surface of liquid He at sufficiently low temperatures
when $E_C$ dominates the kinetic (thermal) energy \cite{Grimes.PRL.42.795, Fisher.PRL.42.798}.
At higher densities, the Fermi energy becomes large
and plays the role of the kinetic energy;
in this case a \textit{quantum} WC can be stabilized
if $E_C$ is much larger than the Fermi energy \cite{Yoon.PRL.82.1744}
and temperature is sufficiently low.
The addition of a strong, quantizing, perpendicular magnetic field
facilitates the formation of a quantum WC as it quenches the kinetic energy
by forcing the electrons into the lowest Landau level (LL)
\cite{Lozovik.JETP.22.11, Lam.PRB.30.473, Levesque.PRB.30.1056}.
Such a magnetic-field-induced WC has been studied using various experimental techniques
in very high mobility (low-disorder) 2DESs confined to modulation-doped GaAs quantum wells
\cite{Andrei.PRL.60.2765, Jiang.PRL.65.633, Goldman.PRL.65.2189, Li.PRL.67.1630, Ye.PRL.89.176802,
Chen.NatPhys.2.245, Tiemann.NatPhy.10.9.648, Hao.PRL.117.096601, MShayegan.Review}.
The measurements have established that,
at LL filling factors $\nu$ smaller than $\simeq 0.20$,
there is an insulating phase
which is generally interpreted to signal the formation of a WC
pinned by the small but ubiquitous disorder potential.
In very low-disorder GaAs 2DESs,
the WC correlation lengths deduced from the measurements
are typically much larger than the WC lattice constant,
implying large domains and long-range spatial order \cite{Ye.PRL.89.176802}.

Here we address a general and fundamental question:
What happens to the WC if one brings a second layer in close proximity;
in particular, how does such a layer modify the Coulomb interaction in the WC layer?
In the case of a classical 2D WC,
theory \cite{Peeter.PRL.50.2021, Peeter.PRB.30.159} and experiments \cite{Mistura.PRB.56.8360}
indicate that placing a conductive plate below the thin liquid He film on which the 2D WC is formed
screens the Coulomb interaction and weakens the stability of the WC.
In order to boost the ratio of the Coulomb to the thermal energy and re-stabilize the WC,
higher electron densities and/or lower temperatures are needed \cite{Peeter.PRL.50.2021, Peeter.PRB.30.159, Mistura.PRB.56.8360}.
The role of screening on a quantum WC, however has not been studied so far.
In our study we probe the stability
of the magnetic-field-induced, quantum WC at very low temperatures
in carefully-designed, bilayer electron systems (BLESs) with very asymmetric layer densities.
The majority-density layer acts as the screening layer,
and influences the critical LL filling factor ($\nu_C$)
below which the WC forms in the minority layer.
Our measured $\nu_C$ in BLESs with different interlayer distances
reveal that, if the screening layer is close by ($\simeq 40$ nm),
$\nu_C$ can be reduced by more than an order of magnitude compared to the $\nu_C \simeq 0.20$ for a single-layer 2DES.
This observation implies that, for a $quantum$ WC,
the weakening of the Coulomb interaction caused by a screening layer
shifts $\nu_C$ (or, equivalently the critical density below which the WC forms) to smaller values.
Our systematic measurements of $\nu_C$
and its dependence on the interlayer distance and other parameters of the BLES
provide unique data which should stimulate quantitative, rigorous calculations.



Our samples are grown via molecular beam epitaxy and contain two 30-nm-wide GaAs quantum wells (QWs),
separated by varying thicknesses of Al$_{0.24}$Ga$_{0.76}$As barriers:
10 nm for samples A and B, 20 nm for sample C, and 50 nm for samples D and E.
The QWs are modulation-doped with Si $\delta$-layers asymmetrically.
As grown, the majority- and minority-layer densities
near zero magnetic field ($B$) are $n_{Maj,0} \simeq 1.45$ and $n_{Min,0} \simeq 0.45$,
in units of $10^{11}$ cm$^{-2}$ which we use throughout the manuscript.
Samples A, C, and D have the majority layer on the top and minority layer on the bottom,
while samples B and E have an inverted layer order.
All the samples have a van der Pauw ($\simeq 4 \times 4$ $mm^2$) geometry,
except for sample A which is a $200 \times 800$ $\mu m^2$ Hall bar.
We use In-Sn alloy to make Ohmic contacts to both layers.
Top and bottom gates are fabricated to tune each layer's density.
In order to measure the longitudinal ($R_{xx}$) and Hall ($R_{xy}$) resistance,
we use low-frequency ($\leq 30$ Hz) lock-in technique and
a dilution refrigerator with a base temperature of $\simeq 30$ mK.


\begin{figure*}[htbp]
\includegraphics[width = 1\textwidth]{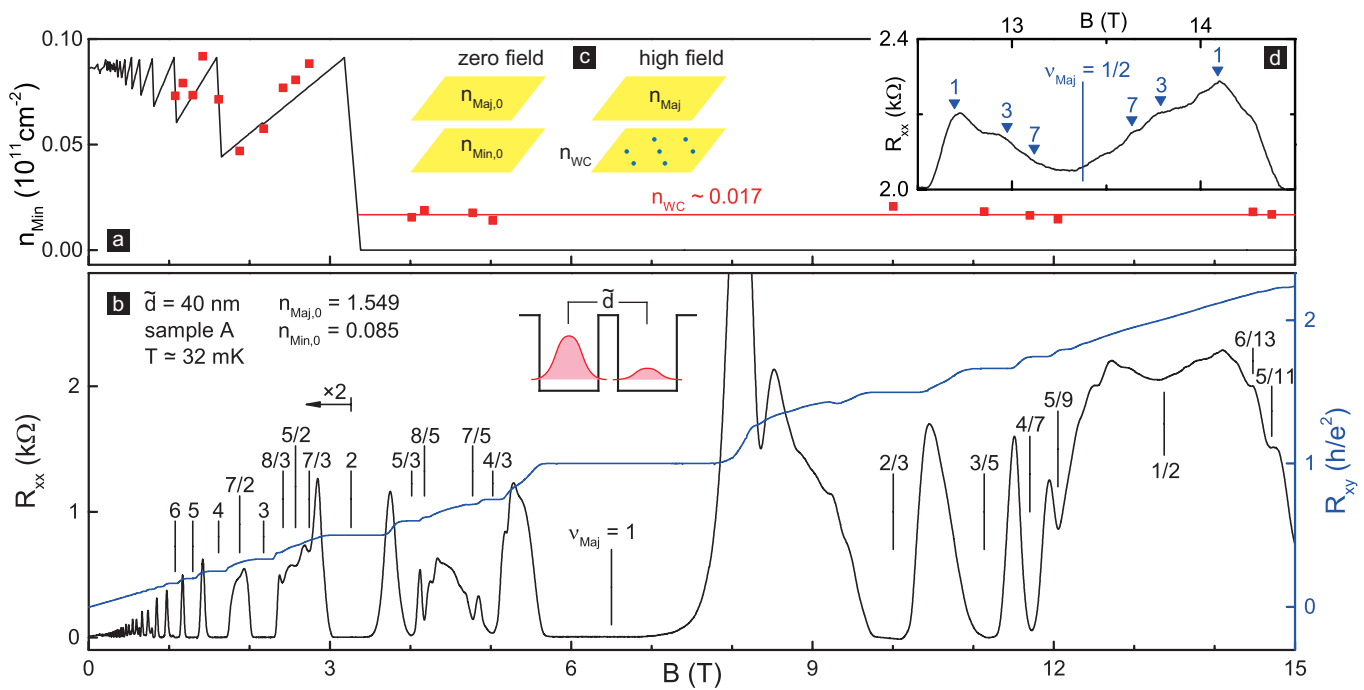}
\caption{
(a) The evolution of minority layer density ($n_{Min}$) with magnetic field for sample A.
Black lines represent the calculated values based on the LLA model
with the zero-field layer densities ($n_{Maj,0}$ and $n_{Min,0}$)
and the interlayer distance as inputs.
Red squares are the measured $n_{Min}$.
Red horizontal line indicates the WC density ($n_{WC}$)
determined by averaging the measured $n_{Min}$ at high $B$ ($ > 3.3$ T).
(b) Magnetotransport traces for sample A: $R_{xx}$ (black), and $R_{xy}$ (blue).
The filling factors for the majority layer ($\nu_{Maj}$) are noted
for the main QHSs and the half fillings.
The $R_{xx}$ trace for $B < 3.3$ T is amplified by a factor of 2 for clarity.
The inset is a diagram for the double-QW sample structure,
showing the definition of the center-to-center distance ($\widetilde{d}$).
(c) Diagram showing the state of each layer at $B = 0$ and high $B$.
(d) Details of $R_{xx}$ near $\nu_{Maj} = 1/2$.
Blue triangles indicate the expected positions of the commensurability maxima
based on the measured $n_{WC}$.
The numbers above the triangles indicate the number of WC lattice points
encircled by the composite fermions' cyclotron orbits \cite{Hao.PRL.117.096601}.
}
\end{figure*}

In Fig. 1, we demonstrate our determination
of the WC density ($n_{WC}$) and $\nu_{C}$ using data for sample A.
Near $B = 0$, we measure $n_{Maj,0}$ and $n_{Min,0}$
from the Fourier transform of the low-field Shubnikov-de Haas oscillations.
The sum of $n_{Maj,0}$ and $n_{Min,0}$ gives the total density,
which we assume remains constant as a function of $B$ even at the largest $B$.
Using $n_{Maj,0}$, $n_{Min,0}$, and the center-to-center distance $\widetilde{d}$
(the sum of the barrier thickness and QW width; see Fig. 1(b) inset) as inputs,
we apply a Landau level alignment (LLA) model to calculate
the evolution of the layer densities as a function of $B$ \cite{Hao.PRB.96.081102, Suppl.Mat}.
The LLA model considers the fact that
applying $B$ induces LLs in each layer,
and that thermal equilibrium requires the Fermi levels
to be the same in both layers.
As a consequence, the evolution of the LLs in each layer
changes the interlayer potential,
and induces an interlayer charge transfer \cite{Hao.PRB.96.081102, Suppl.Mat}.

We plot in Fig. 1(a) (black curve) the calculated $B$-dependence
of the minority-layer density ($n_{Min}$) based on the LLA model.
Experimentally, we measure the majority-layer density ($n_{Maj}$) as a function of $B$
from the positions of the quantum Hall states (QHSs),
namely the minima of the $R_{xx}$ trace in Fig. 1(b) \cite{Footnote_Fig1}.
Because of the much higher $n_{Maj}$,
the $R_{xx}$ minima at intermediate and high $B$
reflect the QHSs of the majority layer \cite{Hao.PRL.117.096601, Hao.PRB.96.081102}.
By subtracting $n_{Maj}$ from the total density,
we deduce $n_{Min}$ as a function of $B$ and plot the data as red squares in Fig. 1(a).
The calculated $n_{Min}$ matches the experimental values reasonably well
in the intermediate $B$ regime where the majority-layer filling factor $\nu_{Maj} > 2$,
but at higher $B$ where $\nu_{Maj} < 2$,
there is a noticeable difference between the calculated and experimental values.
The calculation predicts the complete depletion
of the minority layer following a large charge transfer at $\nu_{Maj} = 2$.
However, the measured $n_{Min}$ attains a finite, constant value up to the highest $B$
[red horizontal line in Fig. 1(a)].

We attribute the residual $n_{Min}$ at high $B$
to the formation of a WC in the minority layer.
Starting with certain $n_{Maj,0}$ and $n_{Min,0}$ [Fig. 1(c)],
as a function of $B$ electrons freely transfer between the layers,
consistent with the LLA model.
At high $B$ ($>3.3$ T), when $n_{Min}$ is sufficiently low
so that the minority-layer filling factor ($\nu_{Min}$) is very small,
the WC in the minority layer becomes energetically favored
and terminates the charge transfer
because a pinned WC is essentially incompressible.
Note that $n_{Min}$ remains constant for $B > 3.3$ T,
consistent with the formation of an incompressible WC \cite{Footnote_comp}.
This is further corroborated
by our observation
of commensurability oscillations (COs)
which support the existence of the WC [Fig. 1(d)].
Near $\nu_{Maj} = 1/2$, the majority-layer electrons form composite fermions (CFs)
and execute cyclotron motion in the effective magnetic field
\cite{Jain.PRL.63.199, Halperin.PRB.47.7312, CFbook.Jain}.
If the CFs feel a periodic electric potential modulation
from a WC layer in close proximity,
$R_{xx}$ exhibits maxima whenever the CFs' cyclotron orbits
encircle a certain integer number of the WC lattice points \cite{Hao.PRL.117.096601}.
In Fig. 1(d), the blue triangles mark the expected positions of COs
based on the modulation from the WC with the density $n_{WC} \simeq 0.017$.
The reasonable agreement between the expected and measured positions of $R_{xx}$ maxima
supports the existence of the WC.

\begin{figure}[tbp]
\includegraphics[width = 0.48\textwidth]{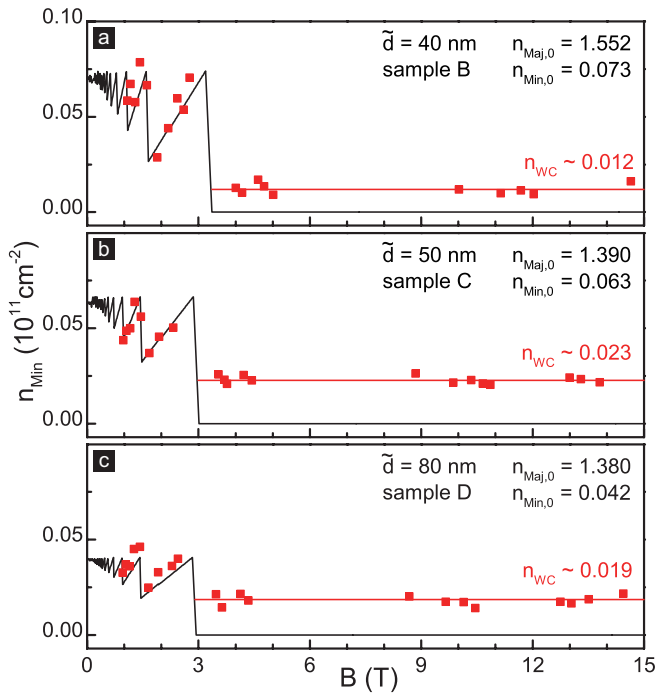}
\caption{
(a)-(c) The evolution of $n_{Min}$ with $B$ for samples with
$\widetilde{d} = 40$, $50$ and $80$ nm.
In each panel,
the black curve is the calculated value of $n_{Min}$,
red squares are the measured $n_{Min}$,
and red horizontal line indicates the value of $n_{WC}$.
}
\end{figure}

The phenomena observed in Fig. 1
is qualitatively seen in our other samples, as illustrated in Fig. 2.
Here we show the calculated and measured values of $n_{Min}$
for samples B, C and, D,
which have $\widetilde{d}$ equal to 40, 50, and 80 nm, respectively.
In all three panels, we start with a sufficiently low $n_{Min,0}$,
and the calculation based on the LLA model predicts a complete depletion
of the minority layer for \textit{all} the samples
at high $B$ when $\nu_{Maj} < 2$.
However, the measurements show that
the minority layer retains a fraction of electrons
in the high-$B$ regime.
The data indicate that the retainment of the electrons
in the minority layer at high $B$ is a general property of the asymmetric BLESs.
As described in the preceding paragraph,
we attribute the retained electrons to the formation of a WC
with density $n_{WC}$ in the minority layer.

\begin{figure*}[htbp]
\includegraphics[width = 1\textwidth]{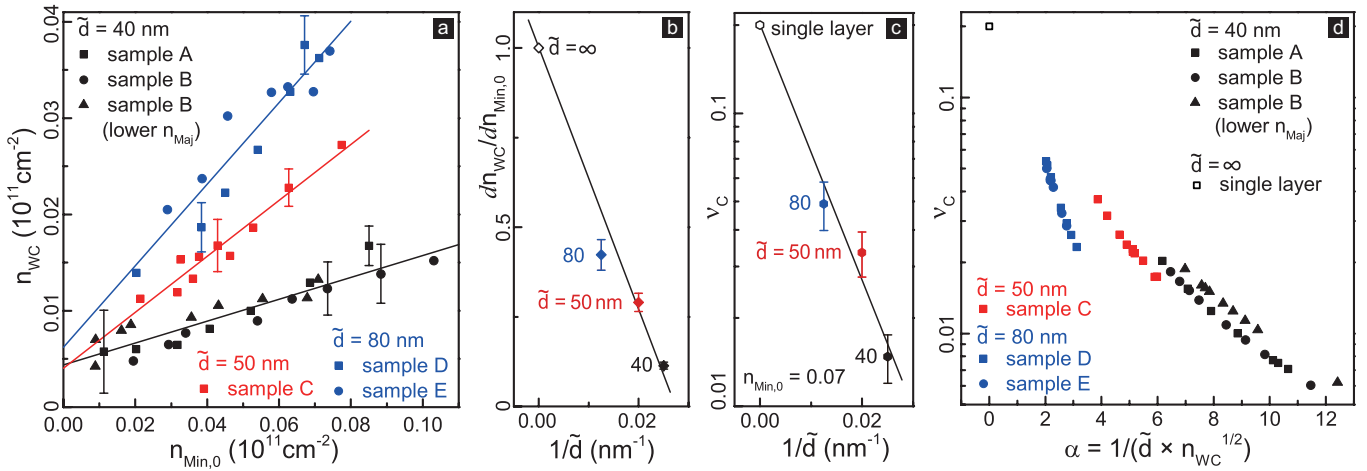}
\caption{
(a) The measured $n_{WC}$ as a function of $n_{Min,0}$.
The black, red, and blue points are data
from samples with $\widetilde{d} = 40$, $50$ and $80$ nm, respectively.
Typical error bars are given,
and the straight lines are the least-squares linear fits for different groups of data.
(b) The slopes of the linear fits in (a) as a function of $1/\widetilde{d}$.
The expected limit of $dn_{WC}/dn_{Min,0}$ at $\widetilde{d} = \infty$
is equal to unity and is plotted by an open symbol.
The black line is a guide to eye.
(c) $\nu_{C}$ as a function of $1/\widetilde{d}$
at fixed $n_{Min,0} = 0.07$.
The error bars include the uncertainty of both $n_{WC}$
and the field position above which the charge transfer stops near $\nu_{Maj} = 2$ in Figs. 1(a) and 2].
The limit of $\nu_{C}$ for the WC in single-layer 2DES (0.20)
is plotted at $1/\widetilde{d} = 0$ as a reference.
The line is drawn to guide the eye.
(d) $\nu_{C}$ as a function of a dimensionless parameter $\alpha$.
The limit of $\nu_{C}$ for the WC in a single-layer 2DES (0.20)
is plotted at $\alpha = 0$ as a reference.
}
\end{figure*}

The value of $n_{WC}$ for a given initial $n_{Min,0}$
directly reflects the impact of the adjacent, majority layer.
In Fig. 3(a), we plot the measured $n_{WC}$ against $n_{Min,0}$ for all the samples.
For each sample, we make measurements for multiple values of $n_{Min,0}$
which we tune by applying voltage bias to the gate
near the minority layer.
Despite the scatter and the measurement error bars,
the data in Fig. 3(a) reveal a clear dependence on $\widetilde{d}$:
For a fixed $n_{Min,0}$, $n_{WC}$ increases with increasing $\widetilde{d}$.
Moreover, for a given $\widetilde{d}$, $n_{WC}$ shows an approximately linear dependence on $n_{Min,0}$.
We would like to point out that
$n_{Maj,0}$ varies slightly for different samples
even though the wafers were designed to have the same density.
Moreover, because of negative compressibility
\cite{Eisenstein.PRB.50.1760, Katayama.SurfSci.305.405, Ying.PRB.52.R11611, Papadakis.PRB.55.9294},
$n_{Maj,0}$ also changes when we use the gate bias to tune $n_{Min,0}$.
In the range of our experiments,
the overall variation of $n_{Maj}$, is $\simeq 10\%$ (see, e.g. $n_{Maj,0}$ listed in Figs. 1 and 2).
To appraise the role of $n_{Maj,0}$,
we also made a full batch of measurements on sample B ($\widetilde{d} = 40$ nm)
with $\simeq 20\%$ lower $n_{Maj,0}$ by applying bias to the gate near the majority layer;
the results are plotted as black triangles in Fig. 3(a).
As seen in Fig. 3(a), the triangles lie only very slightly above the other black data points,
suggesting a weak dependence on $n_{Maj,0}$;
we will return to the role of $n_{Maj,0}$ later in the manuscript.

To further summarize the dependence on $\widetilde{d}$ demonstrated in Fig. 3(a),
we apply a least-squares linear fit to each group of data
belonging to different $\widetilde{d}$,
and plot the slopes ($dn_{WC}/dn_{Min,0}$) of the fitted lines
against $1/\widetilde{d}$ in Fig. 3(b) \cite{Footnote_LinFit}.
Moreover, for each data point in Fig. 3(a), we use $n_{WC}$ and
the $B$ value at $\nu_{Maj} = 2$,
above which the interlayer charge transfer stops,
to determine the $\nu_{C}$ for the formation of the WC.
Figure 3(d) contains a plot of $\nu_{C}$ from all our measurements.
In Fig. 3(c), we show a subset of this data,
namely $\nu_{C}$ for a fixed $n_{Min,0} = 0.07$,
plotted as a function of $1/\widetilde{d}$ \cite{Footnote_Fig3c}.

Figure 3(b) reveals the tendency that, the larger $\widetilde{d}$ is,
the larger the fraction of electrons that stay in the minority layer and form a WC.
At the same time, Fig. 3(c) demonstrates that,
the larger $\widetilde{d}$ is, the larger $\nu_C$ of the minority-layer WC.
Both figures help portray the impact of the majority layer
on the WC formed in the minority layer.
In the case of small $\widetilde{d}$
[e.g., $\widetilde{d} = 40$ nm in Figs. 3(b) and 3(c)],
the nearby majority layer significantly screens the Coulomb interaction
between the minority-layer electrons.
With a weaker Coulomb interaction,
the minority-layer electrons need a smaller $\nu_C$ for the WC
to achieve the dominance of $E_C$ over the kinetic energy.
As a consequence, a smaller fraction of electrons are retained as WC
in the minority layer.
In the case of large $\widetilde{d}$
[e.g., $\widetilde{d} = 50$ and 80 nm in Figs. 3(b) and 3(c)],
the screening by the majority layer becomes less significant,
and the Coulomb interaction between minority-layer electrons
becomes stronger,
resulting in a larger $\nu_C$ and larger fraction of retained electrons.
Note that, in the limit of infinite $\widetilde{d}$,
the minority layer is in effect a single-layer 2DES,
and all the minority-layer electrons should remain in this layer
and form a WC at $\nu_C \simeq 0.2$ [open symbols in Figs. 3(b) and (c)].
The data in Figs. 3(b) and (c) indeed show an asymptotic behavior
toward the expected value in the limit of infinite $\widetilde{d}$.


In Fig. 3(d), we plot $\nu_{C}$ from all the measurements
against a dimensionless parameter $\alpha = 1/(\widetilde{d} \times n_{WC}^{1/2})$
in an attempt to quantify the impact of the screening
by the adjacent majority layer.
Intuitively, the smaller $\widetilde{d}$ is, the stronger the screening,
leading to a smaller $\nu_{C}$ for the formation of WC
because the intralayer Coulomb interaction in the minority layer is further weakened.
Considering that the intralayer Coulomb interaction in the minority layer
is determined by the electrons' average distance $r \propto 1/n_{WC}^{1/2}$,
we compare $\widetilde{d}$ and $r$,
and use their ratio ($\alpha = r/\widetilde{d}$)
to characterize the screening.
In Fig. 3(d), we also plot $\nu_{C} \simeq 0.20$ for the WC
in a single-layer 2DES at $\alpha = 0$ as a reference.
Overall, the data from all the samples with varying $\widetilde{d}$ and $n_{WC}$
generally follow the same tendency toward the limit of single-layer 2DES,
indicating the effect of screening on the formation of WC.

The data of sample B with $\simeq 20\%$ lower $n_{Maj,0}$
[black triangles in Fig. 3(d), $n_{Maj,0} \simeq 1.30$]
show slightly higher $\nu_C$ compared to the data of sample B
with larger $n_{Maj,0}$ (black circles, $n_{Maj,0} \simeq 1.56$).
According to the Thomas-Fermi approximation,
the screening efficiency in a 2DES is independent of the layer density \cite{LDS.Davies},
so $\nu_C$ of sample B should be the same for different $n_{Maj,0}$.
However, previous studies of negative compressibility demonstrate that,
in an interacting BLES, the screening of one layer by another depends on the layer densities
\cite{Eisenstein.PRB.50.1760, Katayama.SurfSci.305.405,
Ying.PRB.52.R11611, Papadakis.PRB.55.9294}.
This interaction-induced, density-dependent screening
beyond the Thomas-Fermi approximation might be responsible for
our observation of the dependence of $\nu_C$ on $n_{Maj,0}$.

In conclusion, our measurements in multiple, asymmetric BLESs
with small minority-layer densities reveal that
the formation of a magnetic-field induced WC at high magnetic fields
retains electrons in this layer and terminates the interlayer charge transfer.
Moreover, we find that the critical filling factor $\nu_{C}$ for WC formation
strongly depends on the interlayer distance
and is significantly lower than $\nu_{C}$ for a single-layer WC,
reflecting the interlayer interaction and screening from the adjacent, majority layer.
We hope that our systematic and quantitative data, summarized in Fig. 3,
would inspire rigorous theoretical work on the physics
of a quantum WC under the impact of screening by a nearby layer.

\begin{acknowledgments}
We thank R. N. Bhatt, L. W. Engel and J. K. Jain for helpful discussions.
We acknowledge the National Science Foundation (Grant DMR 1709076) for
measurements and the Gordon and Betty Moore Foundation
(Grant GBMF4420), the Department of Energy Basic Energy Sciences (Grant
DE-FG02-00-ER45841), and the National Science Foundation (Grants MRSEC DMR 1420541 and
ECCS 1508925) for sample fabrication.
\end{acknowledgments}

\bibliographystyle{h-physrev}

\begin{thebibliography}{10}
\bibitem{Footnote_Fig1}
In order to show QHSs in the low-field regime ($B < 3.3$ T) clearly in Fig.
  1(b), we present the $R_{xx}$ trace from another high-quality sample in van
  der Pauw geometry, which has the same $n_{Maj,0}$, $n_{Min,0}$, and
  $\widetilde{d}$.
\bibitem{Footnote_LinFit}
Because the exact form for the dependence of $n_{WC}$ on $n_{Min,0}$ is
  unknown, we use the linear fit as a simple approximation. The positive
  intercepts of the linear fits might originate from the deviation between the
  linear approximation and the exact form, or the error bar of the experimental
  data points, or both.
\end{thebibliography}

\end{document}